# Energy-efficient Traffic Bypassing in LTE HetNets with Mobile Relays

George T. Karetsos
Dept. of Computer Engineering,
Technology Education Institute of Thessaly, Larissa
GR-41110, Greece
karetsos@teilar.gr

Angelos Rouskas
Dept. of Digital Systems,
University of Piraeus, Piraeus
GR-18534, Greece
arouskas@unipi.gr

Fotis Foukalas
Dept. of Electrical Engineering,
Qatar University, Qatar
foukalas@qu.edu.qa

*Abstract*— One of the core technologies being standardized by 3GPP for LTE-A is the introduction of Relay Nodes (RNs). RNs are intended for ensuring coverage at cell edges as well as for the provision of enhanced capacity at hot spot areas. An extension to this concept is the Mobile Relay (MR). MRs can be mounted on vehicles and the original idea is to serve users inside high speed trains thus counter fighting the inherent severe fading and vehicle penetration loss. In this work we present a framework for exploiting Mobile Relay (MRs) even at low speeds in urban environments for bypassing traffic from nearby users, either within or outside the vehicles. In particular we show that apart from increased capacity and good quality coverage this approach achieves important energy savings for the mobile terminals.

## I. INTRODUCTION

Traffic generated from mobile users is increasing at a very rapid pace. Two major problems associated with this fact is the capability of the wireless networks to serve such a demand on one hand and the energy requirements posed on the terminals that are struggling to transmit and receive huge amounts of data with the limited charge of their batteries one the other hand. Relay nodes (RNs) have been specified in 3GPP Release 9 as a means for extending or ensuring coverage at cell edges, for supporting high data rates, as well as for group mobility and even temporary network deployment [1]. Furthermore in release 11 the concept of Mobile Relays (MRs) is being introduced aiming mainly in providing connectivity with quality for users in high speed vehicles [2]. In particular the scenario that is being considered is for providing good quality access to users within high speed trains. Since MRs present the potential to offer important improvements on the network's operation as well as on the experience received by the users are still under consideration in release 12 [3].

In this work we present a different usage scenario for MRs compared to the one specified in 3GPP related documents [2,3]. In particular MRs can be exploited not only in high speed trains but in other vehicles as well. For example in city roads cars and buses are travelling at relatively low speeds and thus they can become MRs not only for their passengers but also for the pedestrians walking on the pavement beside or for users located at nearby shops and cafes. It is obvious that in such a scenario the users that are moving within or very close to the vehicles that are carrying MRs will enjoy better quality of service when compared with the one received when they are attached to macro cells. This is mainly due to the fact that the distance between the transceivers is quite smaller and thus the users will enjoy better signal quality leading to increased data rates. This fact has been corroborated by the authors in [4]. Furthermore the proposed approach capitalizes also on the well-known fact that mobility increases capacity [5].

We use the term "bypassing" to distinguish from offloading since the traffic does not leave the original network. Bypassing traffic via MRs presents certain advantages when compared with offloading via WiFi that has been the subject of many recent studies [6, 7]. The most important advantage is the fact that all the data transferring operations will remain under the control of a single network operator thus offering increased availability and reliability to the end users. Furthermore the throughput offered nowadays by 4G networks exceeds the one offered by WiFi. This has been revealed via real measurements also. For example the authors in [8] calculated the median throughput of LTE to be three times higher than the one achieved from WiFi at the downlink and six times higher at the uplink. These outcomes are further confirmed in [9]. Furthermore the authors in [10] highlight that the energy efficiency of LTE exceeds that of WiFi and they claim that this will be the case also in the future.

An important aspect in the scenario of using MRs for bypassing users' traffic is the energy savings that can be achieved for the batteries of the mobile terminals. Handheld devices are nowadays used mostly for data transfers and we witness an unprecedented increase on the volumes exchanged. However this comes at the cost of the energy used that puts the batteries of the mobile terminals to their limits. Energy consumption has become one of the major problems faced not only by the users of smartphones and similar devices but also for the respective manufacturers who are striving to cope technically with such increased energy demands.

Energy savings through the exploitation of MRs seems to be almost certain for the terminals that are moving with the vehicle that carry them. However their existence may have a positive impact on the users located on the adjacent pavements or nearby shops if we consider MRs that will be moving mostly within city roads and hot spots in particular. In this case the vehicles will be moving at relatively low speeds and the distance, which is the main factor affecting energy consumption, between the communicating parts will remain small for a duration that will allow a user to use it. However the efficiency of such an operation depends heavily on the user's type of service and on his willingness to accept some delay if at all.

The volume of data generated by smartphones is increasing at a rapid pace due to their capabilities in capturing high definition video and photos. Since these data, most of times, need to be communicated, the associated traffic increases as well. In addition, large amounts of synchronization data are exchanged in the background between the UEs and the networking infrastructure. Such kind of data transfers are not delay sensitive. Even video streaming, which is a bandwidth demanding service, is shown that it can exhibit acceptable performance over delay tolerant networks [11]. In this work we consider data only transfers from applications that produce or handle relatively large amounts of information and are delay tolerant. Delay tolerance provides a degree of freedom to transmit the associated data when conditions are most favorable. In such a case considerable energy savings are expected.

The scope of this work is to examine and confirm first of all the conditions under which it is preferable for a UE to use a passing by MR instead of macro or even picocells. Then we study the energy savings that are possible with this approach and the energy-delay trade off if the users are willing to accept delayed transfers. To the best of our knowledge such a study has not been carried out yet. In particular our contribution is summarized as follows: a) we introduce the concept of traffic bypassing in cellular networks using MRs, b) optimal deployment choices are provided and c) the energy efficiency of the proposed approach as well as its bypassing efficiency for delayed transfers is evaluated via simulations.

This work is organized as follows. In section II the system model is introduced and the main assumptions taken into account are defined. Then in section III we provide simulation results that illustrate the performance of the proposed scheme. Finally conclusions and future work items are highlighted in section IV.

## II. SYSTEM MODEL AND ASSUMPTIONS

We consider an urban environment with picocell deployment along the macrocells forming a heterogeneous networking environment (HetNet) [12]. The picocells are positioned at street junctions and have a transmission range of about 150m [13]. The introduction of MRs can lower this distance to about 30m. As we pointed out already, in this work we focus on users that are located outside the MRs. The reference area and the system model representing our scenario is depicted in fig. 1. A heterogeneous networking environment is considered with both macro and pico eNodeBs deployment and MRs on the streets. Furthermore we assume that the spectrum used on all transmission and reception links belongs to the same operator.

One of the most important factors that defines the efficiency of an MR is its speed since it actually specifies the time window when the communication range will be short and the maximization of the possibility for energy savings. Speed limit in urban roads can go up to 50km/h for cars but recent studies suggest a 30km/h, since such a speed presents the lower fatality risk [14].

Given these facts in this work we assume a median speed of 30km/h which is a valid choice if we consider that the lower speeds can be quite slow or even zero for several seconds when traffic jams are experienced. Furthermore the street lane width is important here since it defines the distance between the user equipment (UE) and the MR. According to several reports and design guidelines, urban city roads have lanes with width ranging between 2.7m and 3.7m and we may have up to 4 such lanes per road [15]. So the mean width is 3.2m, which we adopt.

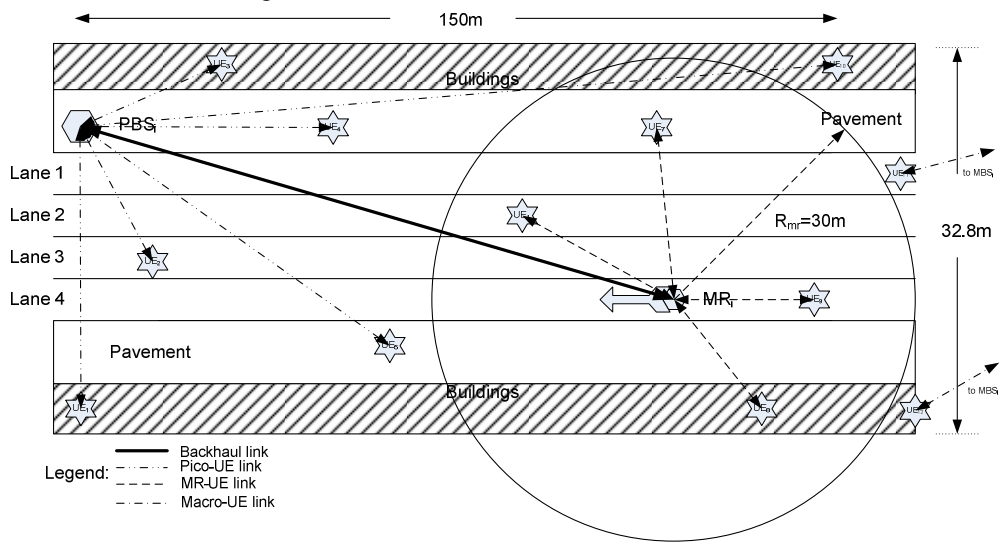

Fig. 1. Reference area and system model

Based on the values presented above we can derive the minimum time window within which it will be possible for a UE to be attached to a passing MR. The layout of the scenario that presents the exploitation of MRs in urban roads for bypassing traffic from adjacent UEs is depicted in fig. 1. In order to validate that there exists enough time for data such an operation we assume the worst case situation which is when the MR is travelling in the lane that is located further from the UE. Taking into account that the MR antennas will be positioned on the middle of the lane and assuming a pavement of 5m and another 5m for in building penetration the maximum possible radio distance between the MR and a UE is calculated at 21.2m. This means that the distance the MR will travel while providing coverage to the UE will be 42.4m as depicted in fig. 1. If we adopt a 50km/h travelling speed for the MR the minimum coverage time will be 3.05sec.

In these calculations we should also take into account the time required for the handover to be executed. Considering that both the picocell and the MR are E-UTRAN based, the handover delay according to the 3GPP guidelines is 50ms if the target cell is known or 130ms for an unknown target cell [16]. Furthermore a recent study that is based on measurements indicates a mean handover time over the X2 interface of 30ms while the data interruption time was found to have a mean value of 50ms [17]. Thus we can assume that the handover procedure has minimal impact on the overall coverage time and especially for the type of services targeted in this study.

LTE aims by design to achieve very high throughputs. The targeted peak data rate values according to 3GPP Release 8 are 300Mbps for the downlink and 75Mbps for the uplink. However recent studies based on measurements report much lower peak data rates for both uplink i.e. from the UE to the MR and downlink i.e. from the MR to the UE. In particular [18] reports a peak data rate of 16.4Mbps for the uplink and 61.2Mbps for the downlink with 20MHz channel bandwidth. Furthermore the median uplink and downlink throughput achieved by LTE for different users in different locations is 5.64Mbps and 12.74Mbps respectively and has been calculated via extensive measurements that are documented in [8]. Thus even if we take the worst case situation, at least 1.8Mbytes uplink and 4.3Mbytes downlink could be transferred during a single, one second attachment of a UE to an MR. This outcome can lead to important energy savings for both the UE as well as for the infrastructure. In this work our focus is the detailed study of energy savings at the UEs only and thus on uplink transmissions from the UE to the MR. A detailed assessment of the performance gains of such a scenario that targets downlink transmissions is left for a future study.

*A. UE-MR attachment procedure*

The attachment of a UE to an MR is based on the Reference Signal Received Power (RSRP). When an $MR_i$ is passing, the adjacent UEs will attach to it due to the fact that the received RSRP is higher from that received from the pico or macro base station (PBS or MBS respectively). However the attachment request will be fulfilled only if the speed $u_{mri}$ of the $MR_i$ is less than a predefined threshold $u_{mrt}$ that will allow for sufficient data to be transferred, and that there is enough capacity $C_{mrt}$ available. Furthermore no voice call should be ongoing. Finally the distance of the MR from the pico BS should not be less than $d_{mrp}$. In order for the latter to be accomplished the MRs are GPS equipped and know the positions of the network's BSs in the area under consideration. This information is updated periodically. The complete procedure is provided in Algorithm 1 using pseudo code.

---

**Algorithm 1:** Handling of $UE_i$ attach request to $MR_i$

```
function attach(UEi, MRi);
Input: Request from UEi to MBSi or PBSi to
       attach to MRi
Output: UEi attached to MRi or continues with
        MBSi or PBSi
UEi attach request to MRi
if UEi attached to MBSi then
   if service type is data then
      if umri≤umrt then
         if Cmri≥ Cmrt then
UEi attach request to MRi granted
else
UEi continues with MBSi
         end if
      end if
   end if
else if UEi attached to PBSi then
   if service type is data then
      if umri≤umrt then
         if Cmri≥ Cmrt then
            if d>dmrp then
UEi attach request to MRi granted
else
UEi continues with PBSi
            end if
         end if
      end if
   end if
end if
end
```

---

### III. RESULTS AND DISCUSSION

In this section we present simulation results regarding bypassing efficiency and energy efficiency of the proposed approach that were derived using Matlab. First of all and based on the layout presented in fig. 1 we estimate the traffic density in the area under consideration. Since the area we target is a mixture of urban and dense urban environment a reasonable value for the number of active users is $u_a$=25 given the fact that in urban environments we may have around $10^3$ users per km$^2$ while in dense urban we may have around $10^4$ users per km$^2$. Since the coverage area of the MR is $A_{MR}$=2826m$^2$ while the total test area is $A_{tot}$=4920m$^2$ we can assume $u_{MR}$=10, $u_{PBS}$=10 and $u_{MBS}$=5 users respectively. Using the results presented in [19] that correspond only to data traffic and the fact that the downlink to uplink traffic ratio is around 6:1 [20], the estimated traffic density at the uplink for the given area is $tr_{Dtot}$=4Mbytes. Thus the traffic density per user at the uplink at a given moment in time during the busy hour (BH) is $tr_{Du}= tr_{Dtot}/u_a$=0.16Mbytes if we consider a uniform share that is acceptable in the adopted scenario. Furthermore, for simplicity, we assume the full buffer traffic model i.e. each UE has always data ready for transmission and no mobility. We would like to note here that the study presented in [20] was done back in 2010 and the ratio presented is changing in favor of the uplink due to the advanced capabilities of the newest smartphones in taking photos and videos.

In fig. 2 we present the average attachment time of a UE to an MR in relation to the speed of the passing MR assuming that the test area belongs to a macrocell only or it is a HetNet with both macrocell and picocell coverage. Obviously the attachment time is less when the speed of the MR is increasing. Most gains are expected when MRs' speed is less than 30km/h which surprisingly corresponds to the suggested speed limit for urban environments [8]. We also observe that the attachment time is less when the area belongs to a HetNet environment since in such a situation the criterion $d>d_{mrp}$ of Algorithm 1 will not always hold and the UE will remain attached to the respective $PBS_i$.

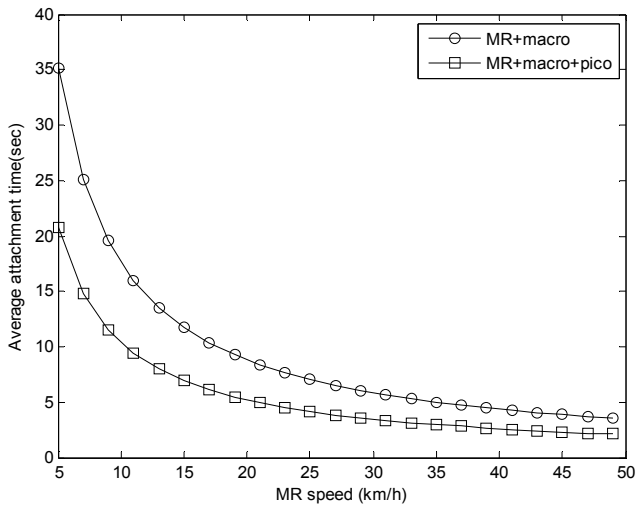

Fig. 2. Average user attachment time vs. MR speed

Fig. 3 presents the availability ratio of an MR to a UE i.e. the time percentage a UE can be attached to an MR in relation to each speed for various MR interarrival periods. Since our focus is to find the lower extremes of the availability ratio that are worth exploiting we derive results for MR speeds higher than 5km/h. Obviously the availability ratio will increase rapidly for speeds close to 0km/h i.e. for stationary MRs blocked in traffic jams. Furthermore, since most of the times MRs will be deployed on public transportation means, we derive and present results for interarrival periods of 3, 6 and 9 minutes which are reasonable values in urban environments. We stopped the simulation at 9min since over this value the availability ratio is getting very small. Indeed, we observe that if the interarrival period is equal or greater to 9min the availability ratio is below 2% for MR speeds greater than 10km/h. This is an important finding for an operator who wants to deploy MRs in the most efficient way. Thus there is a trade-off between deployment cost, energy savings and perceived quality of experience (QoE) for the end user that governs such a decision.

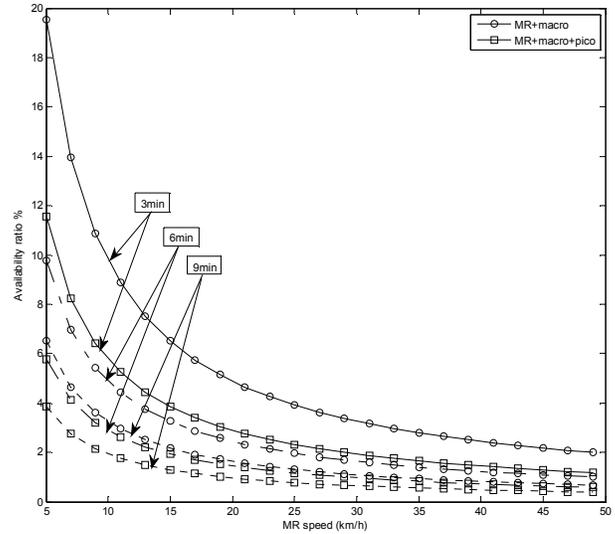

Fig. 3. Availability ratio vs. MR speed for different interarrival times

Given the availability ratio we can estimate the percentage of data that can be bypassed through an MR from a UE during the BH. Considering that a user generates 0.16Mbyte/sec during the BH as pointed out above. we calculate the traffic volume being bypassed in relation to the availability ratio for only macrocell or both macrocell and picocell configurations. The results are presented in fig. 4 and are obtained for a 6 minute interarrival time. Obviously there is an upper limit on the traffic that can be bypassed that is lower in the macro-pico configuration.

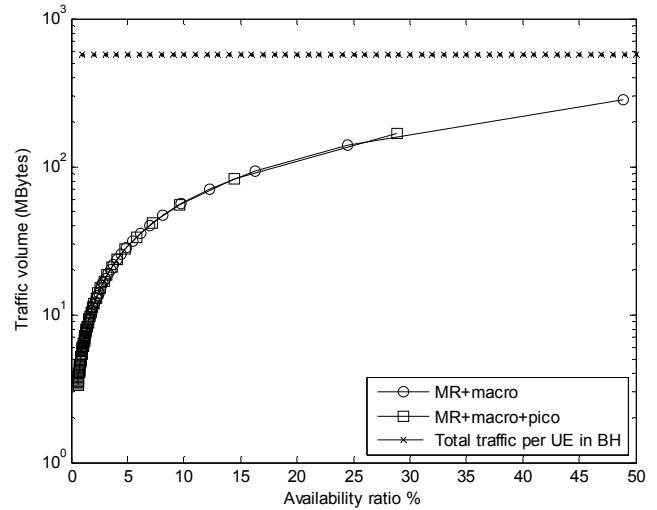

Fig 4. Bypassed traffic vs. availability ratio

In order to calculate the consumed energy required for the transmission of a specific data volume and dimension the energy savings for the presented approach, we adopt an inter-site distance (ISD) of 500m for macrocells as suggested by the 3GPP model given in [1]. This is typical for HetNet LTE deployments in urban environments and corresponds to a cell radius of 289m for the macrocell [21]. Furthermore we assume that all transmissions are taking place in the higher LTE band 7 that operates at 2.6GHz. For the calculations of energy consumption we rely on the results presented in [22]. In fig. 5 we present the energy efficiency of using MRs for bypassing traffic. In particular we compare our scenario i.e. MR plus macro and MR plus pico with only macro or pico configurations. We observe that we achieve gains in both situations. However the energy savings are quite larger when the MRs are enhancing the operations of only macro configurations. Indeed as the availability ratio increases, the performance of this option increases in a ratio quite higher than the MR plus pico configuration. In particular we observe that energy efficiency of the MR plus macro set-up is approaching the MR plus pico one for availability ratios close or larger than 50%. This fact presents an attractive alternative to picocell deployment that could serve as a solution for capacity provisioning and energy savings in an ad-hoc manner.

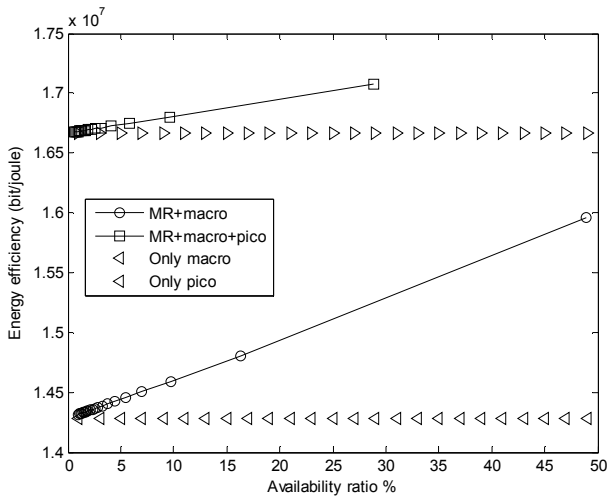

Fig. 5. Energy efficiency vs. availability ratio

Finally in fig. 6 we present the bypassing efficiency of the proposed approach when the users can sustain artificially delayed transfers that can be controlled via the applications on their smartphones. In particular we assume that a user wants to transmit 100Mbytes of data corresponding to background traffic or whatever traffic for which he can accept a delay. Then we compute the bypassing efficiency for various delays. In particular we observe that if the user is willing to accept about 57 minutes of delay 100% of the data could be transmitted via the MR if we assume the very modest value of 1.8Mbyte/s as the median uplink throughput [8]. In the same time about 59% of the data file is bypassed when picocells are deployed. Due to this fact, energy savings will be less in such a configuration when compared with the macro plus MR one. Finally in fig. 6 one can observe that very close to 0 minutes there exists bypassing efficiency. This is because we assume that an MR is already present when we start the simulation.

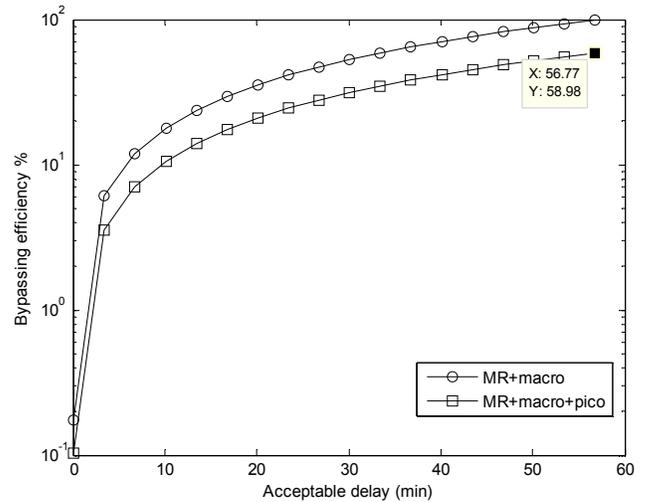

Fig. 6. Bypassing efficiency for delayed transfers

The energy that is saved for bypassing 100Mbytes of data through MRs in the macrocell configuration is about 1.57Joule. This amount is enough for transmitting about 26Mbit of data via a picocell or about 22.5Mbit via a macrocell and highlights the importance of the proposed approach.

## IV. CONCLUSIONS AND FUTURE WORK

In this work we introduced the concept of using mobile relays to bypass data traffic generated by mobile devices in urban environments in order to enhance their energy efficiency and prolong battery life. By studying a particular test area we derive the average attachment time and the availability ratio and we show that for best performance the speed of the MRs should be kept below 30km/h while their interarrival times should be less than 9min. Furthermore we derive the energy efficiency and the bypassing efficiency of the proposed approach. Since this work serves mostly as a proof of concept, there is much space for research that is planned for the future. In particular two major foreseen research items are related with interference assessment and management as well as with the detailed study of the tradeoff between deployment cost vs. energy efficiency, as it has been already pointed out in the results section.

## ACKNOWLEDGMENT

This research has been co-financed by the European Union (European Social Fund - ESF) and Greek national funds through the Operational Program "Education and Lifelong Learning" of the National Strategic Reference Framework (NSRF) - Research Funding Program: ARCHIMEDES III. Investing in knowledge society through the European Social Fund.

REFERENCES

[1] 3GPP, "Technical Specification Group Radio Access Network; Evolved Universal Terrestrial Radio Access (E-UTRA); Further advancements for E-UTRA physical layer aspects (Release 9)", TR 36.814 V9.0.0, March 2010.

[2] 3GPP, Technical Specification Group Radio Access Network; Evolved Universal Terrestrial Radio Access (E-UTRA); Study on mobile relay (Release 11)", TR 36.836 V1.0.0, June 2014.

[3] 3GPP, Technical Specification Group Radio Access Network; Evolved Universal Terrestrial Radio Access (E-UTRA); Study on mobile relay (Release 12)", TR 36.836 V12.0.0, May 2012.

[4] Kokkoniemi, Joonas; Ylitalo, Juha; Luoto, Petri; Scott, Simon; Leinonen, Jouko; Latva-aho, Matti, "Performance evaluation of vehicular LTE mobile relay nodes," *Personal Indoor and Mobile Radio Communications (PIMRC), 2013 IEEE 24th International Symposium on*, vol., no., pp.1972,1976, 8-11 Sept. 2013.

[5] Diggavi, S.N.; Grossglauser, M.; Tse, D.N.C., "Even One-Dimensional Mobility Increases the Capacity of Wireless Networks," *Information Theory, IEEE Transactions on*, vol.51, no.11, pp.3947,3954, Nov. 2005.

[6] Qutqut, M.H.; Al-Turjman, F.M.; Hassanein, H.S., "MFW: Mobile femtocells utilizing WiFi: A data offloading framework for cellular networks using mobile femtocells," *Communications (ICC), 2013 IEEE International Conference on*, vol., no., pp.6427,6431, 9-13 June 2013.

[7] Hinger, D.; Kalbande, D., "Investigation of throughput gains by mobile data offloading from LTE to Wi-Fi," *India Conference (INDICON), 2014 Annual IEEE*, vol., no., pp.1,6, 11-13 Dec. 2014.

[8] J. Huang, F. Qian, A. Gerber, Z. M. Mao, S. Sen, and O. Spatscheck, "A Close Examination of Performance and Power Characteristics of 4G LTE Networks". In MobiSys, 2012.

[9] Shuo Deng, Ravi Netravali, Anirudh Sivaraman, and Hari Balakrishnan, "WiFi, LTE, or Both?: Measuring Multi-Homed Wireless Internet Performance", In *Proceedings of the 2014 Conference on Internet Measurement Conference* (IMC '14). ACM, New York, NY, USA, 181-194.

[10] M. Lauridsen, L. Noël, T.B. Sørensen and P. Mogensen "An *Empirical LTE Smartphone Power Model* with a *View* to *Energy Efficiency Evolution*", Intel® Technology Journal, Volume 18, Issue 1, 2014.

[11] Morgenroth, Johannes, Tobias Pögel, and Lars Wolf, "Live-streaming in delay tolerant networks." *Proceedings of the 6th ACM workshop on Challenged networks*, pp. 67-68, 2011.

[12] Khandekar, A.; Bhushan, N.; Ji Tingfang; Vanghi, V., "LTE-Advanced: Heterogeneous networks," *Wireless Conference (EW), 2010 European*, vol., no., pp.978,982, 12-15 April 2010.

[13] Mellios, E.; Hilton, G.S.; Nix, A.R., "Evaluating the impact of user height variations on Outdoor-to-Indoor propagation in urban macrocells and picocells using ray-tracing," *General Assembly and Scientific Symposium (URSI GASS), 2014 XXXIth URSI*, vol., no., pp.1,4, 16-23 Aug. 2014.

[14] Rosén, E., H. Stigson and U. Sander, "Literature review of pedestrian fatality risk as a function of car impact speed", *Accident Analysis and Prevention* Vol. 43, No. 1, 2011, pp. 25-33.

[15] Hall, L. E., R. D. Powers, D. S. Turner, W. Brilon, and J. W. Hall. "Overview of cross section design elements." In *International Symposium on Highway Geometric Design Practices, Boston*. 1995.

[16] 3GPP, Technical Specification Group Radio Access Network; Evolved Universal Terrestrial Radio Access (E-UTRA); Requirements for support of radio resource management (Release 12), TS 36.133 V12.7.0 March 2015.

[17] Elnashar, A.; El-Saidny, M.A., "Looking at LTE in Practice: A Performance Analysis of the LTE System Based on Field Test Results," *Vehicular Technology Magazine, IEEE*, vol.8, no.3, pp.81,92, Sept. 2013.

[18] Yi-Bing Lin; Pin-Jen Lin; Sung, Y.C.; Yuan-Kai Chen; Whai-En Chen; Alrajeh, N.; Lin, B.-S.P.; Chai-Hien Gan, "Performance measurements of TD-LTE, WiMax and 3G systems," *Wireless Communications, IEEE*, vol.20, no.3, pp.153,160, June 2013.

[19] Dongheon Lee; Sheng Zhou; Xiaofeng Zhong; Zhisheng Niu; Xuan Zhou; Honggang Zhang, "Spatial modeling of the traffic density in cellular networks," *Wireless Communications, IEEE*, vol.21, no.1, pp.80,88, February 2014.

[20] Falaki, H., Lymberopoulos, D., Mahajan, R., Kandula, S., & Estrin, D., "A first look at traffic on smartphones", In *Proceedings of the 10th ACM SIGCOMM conference on Internet measurement,* pp. 281-287, November 2010.

[21] Benjebbour, A.; Anxin Li; Saito, Y.; Kishiyama, Y.; Harada, A.; Nakamura, T., "System-level performance of downlink NOMA for future LTE enhancements," *Globecom Workshops (GC Wkshps), 2013 IEEE*, vol., no., pp.66,70, 9-13 Dec. 2013.

[22] Dusza, B.; Ide, C.; Liang Cheng; Wietfeld, C., "An accurate measurement-based power consumption model for LTE uplink transmissions," *Computer Communications Workshops (INFOCOM WKSHPS), 2013 IEEE Conference on*, vol., no., pp.49,50, 14-19 April 2013.